\documentstyle[aps,preprint,psfig]{revtex}

\begin{document}
\title{Topology of evolving networks: local events and universality}
\author{R\'eka Albert and Albert-L\'aszl\'o Barab\'asi$^*$}
\address{Department of Physics,
        University of Notre-Dame, Notre-Dame, IN 46556}
\maketitle

\begin{abstract}
Networks grow and evolve by local events, such as the
addition of new nodes and links, or rewiring of links from one
node to another. We show that depending on the frequency of these processes two topologically different networks can emerge, the connectivity distribution following  either a generalized power-law or an exponential. We propose a continuum theory that predicts these two regimes as well as the scaling function and the exponents, in good agreement with the numerical results. Finally, we use the obtained predictions to fit the
connectivity distribution of the network describing the
professional links between movie actors.
\end{abstract}
\newpage
The complexity of numerous social, biological or
communication systems is rooted in the rather
interwoven web defined by the system's components and their
interactions. For example, cell functioning is
guaranteed by a complex metabolic network, whose
nodes are substrates and enzymes, and edges represent chemical interactions
\cite{biology}. Similarly, the society is characterized by a huge social network whose nodes are individuals or
organizations, connected by social interactions
\cite{social}, but equally complex networks appear in the
business world, where nodes are companies and edges
represent diverse trade relationships, or describe the
world-wide web (www), whose nodes are HTML documents
connected by links pointing from one page to another
\cite{clever,ibm,diam}.

The study of random networks, a much investigated topic in
the mathematical literature, has been dominated by the model
of Erd\H{o}s and R\'{e}nyi (ER)
\cite{er} that views the network as a set of nodes, each pair of nodes being connected with
equal probability.  Recently Watts and Strogatz (WS)
\cite{ws} found that local clustering is an important
characteristic of random networks, offering the first indication that real
networks could be more complex than predicted by the ER model. A common feature of the ER and WS models is that the probability $P(k)$ that a node in the network is connected to $k$ other nodes is bounded, decaying exponentially for large $k$. In contrast, exploring several large
databases describing the topology of large networks, recently we found
\cite{ba}
that independently of the nature of the system
and the identity of its constituents, $P(k)$ decays as a power-law, following $P(k)
\sim k^{-\gamma}$. The generic nature of this result was supported by measurements on the www connectivity\cite{clever,ibm,diam}, actor networks\cite{ws,ba}, citation network of scientists\cite{redner}, and recent results on the Internet topology\cite{faloutsos}. These results offered the first
evidence that some large networks can self-organize into a
scale-free state, a feature unexpected by all previous network models
\cite{er,ws}. The origin of this scale-free behavior has been traced
back to two mechanisms that are present in many
systems, and have a strong impact on the final topology
\cite{ba}. First, networks develop by the addition of new nodes that are connected
to those already present in the system.  Second,
there is a higher probability that a new node is linked
to a node that already has a large number of
connections. These two ingredients led to the formulation of
the scale-free model that generates a network for which $P(k)$ follows a
power-law with $\gamma=3$. While this model
correctly predicts the emergence of power-law scaling, the
agreement between the measured and predicted exponents is
less than satisfactory: for real systems
$\gamma$ is scattered between $2.1$ and $4$\cite{ba}, raising important questions about the universality of network formation.

In this paper we take an important step towards understanding the effect of various local events on the
large scale topology of a network. We introduce and investigate an
extended model of network evolution that gives a more realistic description of the local processes, incorporating the
addition of new nodes, new links, and the rewiring of
links.
Using a continuum theory we show that, depending on the
relative frequency of these local processes, networks can develop two fundamentally different topologies. In the first regime $P(k)$ has a power-law tail, but the exponent $\gamma$ depends
continuously on the relative frequency of the local events.
In the second regime the power-law scaling breaks down, and $P(k)$ approaches an
exponential.  We derive a phase diagram that predicts the transition between these two topologically distinct regimes, and support the analytical predictions with numerical simulations. Finally, we use the obtained prediction to
successfully fit the connectivity distribution of the actor
network, allowing us to determine the frequency of the local reorganization events.

{\it Extended model}: In the scale-free model introduced in Ref.\cite{ba}, the only process affecting the network's topology was the addition of new nodes. The extended model described below offers a more realistic description of network formation by incorporating additional local events that are known to appear in real networks. We start with $m_0$ isolated nodes,
and at each timestep we perform one of the following three
operations (see Fig.$\,$1a):

(i)  {\it With probability $p$ we add $m$($m\leq m_0$) new
links}. For this we randomly select a node as the starting point of the new
link, describing, for example, that a web
developer decides to add a new
URL link to a page.  The other end of the link, however,
is selected with probability

\begin{equation}
\label{prefer}
\Pi(k_i)=\frac{k_i+1}{\sum_j (k_j+1)},
\end{equation}
incorporating the fact that new links preferentially point to
popular nodes, with a high number of connections
\cite{ba}. This process is repeated $m$ times.

(ii) {\it With probability $q$ we rewire $m$ links}. For this
we randomly select a node $i$ and a link $l_{ij}$ connected to
it. Next we remove this link and replace it with a new link
$l_{ij'}$ that connects $i$ with node $j'$ chosen with
probability $\Pi(k_j')$ given by (\ref{prefer}). This process is repeated $m$ times.

(iii) {\it With probability $1-p-q$ we add a new node}.
The new node has $m$ new links that with probability
$\Pi(k_i)$ are connected to nodes $i$ already present in
the system.

Since our goal is to investigate the generic mechanisms of network evolution, we use bidirectional links. However, our results can be easily generalized to directed networks as well. In the model, the probabilities $p$ and $q$ can be varied in the
interval $0\leq p<1$ and $0\leq q< 1-p$. Note that we choose the probability $\Pi(k_i)$
to be proportional to $k_i+1$, such that there is a
nonzero probability that isolated nodes ($k_i=0$)
acquire new links. 
Finally, in the $p=q=0$ limit the model reduces to the scale-free model investigated in Ref.\cite{ba}.

{\it Continuum Theory}: In the model the probability that a node $i$ increases its connectivity $k_i$ depends only on $k_i$ and quantities characterizing the whole network (the parameters $p$, $q$, $m$, the number of nodes and links). To predict the topology and the
dynamics of the network, we assume that $k_i$ changes continuously, and thus the probability
$\Pi(k_i)$ can be interpreted as the rate at which $k_i$
changes\cite{physica}. Consequently, the processes (i-iii) all contribute to $k_i$, each being incorporated in the continuum
theory as follows:

(i) Addition of $m$ new links with probability $p$:

\begin{equation}
\left(\frac{\partial k_i}{\partial
t}\right) _{({\rm i})}=pA\frac 1 N+pA\frac{k_i+1}{\sum_j
(k_j+1)},
\end{equation}
where $N$ is the size of the system. The first term on the r.h.s. corresponds to the random selection of one end of the new
link, while the second term reflects the preferential
attachment (\ref{prefer}) used to select the other end of the link. Since
the total change in connectivity is $\Delta k=2m$, we have
$A=m$.

(ii) Rewiring of $m$ links with probability $q$:

\begin{equation}
 \left(\frac{\partial k_i}{\partial t}\right)_{({\rm ii})}=-qB\frac
1 N+qB\frac{k_i+1}{\sum_j
(k_j+1)} .
\end{equation}
The first term incorporates the decreasing connectivity
of the node from which the link was removed, and the second term represents the increasing connectivity of the node the link is reconnected to. The total
connectivity does not change during the
rewiring process, but $B$ can be calculated by separating the
two processes, obtaining $B=m$.

(iii) Addition of a new node with
probability $1-p-q$:

\begin{equation}
\left(\frac{\partial
k_i}{\partial t}\right)_{({\rm iii})}=(1-p-q)C\frac{k_i+1}{\sum_j
(k_j+1)}.
\end{equation}
The number of links connecting the new node to
the nodes in the system is $m$, thus we have $C=m$.

Since these three processes take place simultaneously, we
have to sum up their contributions, obtaining the continuum
theory describing the change in $k_i$ for the extended model

\begin{equation}
\label{partial}
\frac{\partial k_i}{\partial t}=(p-q)m\frac
1N+m\frac{k_i+1}{\sum_j (k_j+1)}.
\end{equation}In (5) the system size $N$ and the total number of links $\sum_j k_j$ vary with time as
$N(t)=m_0+(1-p-q)t$ and $\sum_j k_j=(1-q)2mt-m$, indicating that for large $t$ we can neglect the constants
$m_0$ and
$m$ compared to the terms linearly increasing with
time. Using as initial condition the
connectivity of a node added at time $t_i$,
$k_i(t_i)=m$, the solution of (\ref{partial}) for
$k_i(t)$ has the form
\begin{equation}
\label{sol}
k_i(t)=\left(A(p,q,m)+m+1\right)\left(\frac{t}{t_i}\right)^{\frac{1}{B(p,q,m)}}-A(p,q,m)-1,
\end{equation}
where
\begin{eqnarray}
A(p,q,m)&=&(p-q)\left(\frac{2m(1-q)}{1-p-q}+1\right),\nonumber\\
B(p,q,m)&=&\frac{2m(1-q)+1-p-q}{m}.
\end{eqnarray}
The probability
that a node has a connectivity $k_i(t)$ smaller than
$k$, $P(k_i(t)<k)$, can be written as
$
P(k_i(t)<k)=P(t_i>C(p,q,m)t),
$
where 
\begin{equation}
C(p,q,m)=\left(\frac{m+A(p,q,m)+1}{k+A(p,q,m)+1}\right)^{B(p,q,m)}.
\end{equation}
Since $t_i$ must satisfy the condition $0\leq t_i\leq t$,
we can distinguish three cases:

(a) If $C(p,q,m)>1$,
then $P(k_i(t)<k)=0$. Thus the condition that $P(k)$ is nonzero is that $k>m$.

(b) If $C(p,q,m)$ is not real, then $P(k_i(t)<k)$ is not well-defined.
Thus to be able to calculate $P(k)$ we must have $\frac{m+A(p,q,m)+1}{k+A(p,q,m)+1}>0$ for all $k>m$, 
satisfied if $A(p,q,m)+m+1>0$.  

(c) Finally, if $0<C(p,q,m)<1$,
the connectivity distribution $P(k)$ can be determined analytically. Defining the
unit of time in the model as one growth/rewire/new link
attempt, the probability density of
$t_i$ is $P_i(t_i) = 1/(m_0+t)$, thus
\begin{equation}
P(k_i(t)<k)=1-C(p,q,m)\frac{t}{m_0+t}
\end{equation} from which, using
$P(k)=\frac{\partial P(k_i(t)<k)}{\partial k},$ we obtain
\begin{equation}
\label{power}
P(k)=\frac{t}{m_0+t}D(p,q,m)\left(k+A(p,q,m)+1\right
)^{-1-B(p,q,m)},
\end{equation}
where $D(p,q,m)=(m+A(p,q,m)+1)^{B(p,q,m)}B(p,q,m)$.

Thus the connectivity distribution, the main result provided
by the continuum theory, has a generalized power-law form
\begin{equation}
\label{probab}
P(k)\propto(k+\kappa(p,q,m))^{-\gamma(p,q,m)},
\end{equation}
where $\kappa (p,q,m)=A(p,q,m)+1$ and $\gamma(p,q,m)=B(p,q,m)+1$.

{\it Phase diagram and scaling:} Eqs.$\,$(\ref {power}) and
(\ref {probab}) are valid only when $A(p,q,m) + m+1>0$ (see (a) and (b)), which, for fixed $p$ and $m$, translates into
$q<q_{max}={\rm min}\{1-p,(1-p+m)/(1+2m)\}$. The
$(p,q)$ phase diagram, shown in Fig.$\,$1b,
indicates the existence of two regions in the parameter
space: For
$q<q_{max}$ $P(k)$ is given by (\ref{probab}), thus the connectivity distribution is scale-free.  For $q>q_{max}$, however, Eq.(\ref{probab}) is not valid, the
continuum theory failing to predict the behavior of
the system. We will demonstrate that in this regime
$P(k)$ crosses over to an exponential. The boundary between the scale-free
and the exponential regimes depends on the parameter
$m$. In the limit
$m\rightarrow 0$ for any
$p+q<1$ $P(k)$ is scale-free, while in the limit
$m\rightarrow\infty$ we have $q_{max}=0.5$, the boundary approaching
a horizontal line. For finite $m$ the phase boundary is a
line with slope $-m/(1+2m)$ (Fig.$\,$1b).

{\it Scale-free regime:} While a power-law tail is present in any
point of this regime, the scaling is
different from that predicted by the simpler scale-free model\cite{ba}. First, according to (\ref{power}), for small
$k$ the probability saturates at $P(\kappa(p,q,m))$.
This feature reproduces the results obtained for real
networks: $P(k)$ obtained for the actor network or the citation network of scientists saturate at small $k$\cite{ba}. Second, the exponent
$\gamma(p,q,m)$ characterizing the tail of $P(k)$ for
$k>>\kappa(p,q)$ changes continuously with $p$, $q$ and $m$,
predicting a range of exponents between 2 and $\infty$. This allows us to account for the wide variations seen in real networks, for which $\gamma$ varies from
$2$ to $4$. To demonstrate the predictive power of the continuum theory, we have
studied the extended discrete model numerically, as defined
in (i-iii). As Fig.$\,$2a shows, Eq.$\,$(\ref{power}) offers an excellent, parameter free fit to the numerical results.

{\it Exponential regime}: For $q>q_{max}$ the continuum theory
is no longer valid, but we can investigate $P(k)$ using
numerical simulations. As Fig.$\,$2b shows, in this regime, as $q\rightarrow 1$,
$P(k)$ develops an exponential tail. This transition to an exponential can be understood in the terms of Model B discussed in Refs \cite{ba,physica},  demonstrating that growth is an essential condition for power-law scaling. However, in the limit $q\rightarrow 1$ growth is suppressed, the frequent rewiring process acting on a network with an almost constant number of nodes. The convergence of $P(k)$ to an exponential in this regime indicates that for this choice of parameters the model belongs to the class of networks defined by the ER\cite{er} and WS\cite{ws} models\cite{mechanisms}.  

{\it Application to real networks:} To illustrate the predictive power of the obtained  results, we choose to investigate the collaboration graph of movie actors
\cite{ws}, which is a non-directed network, thus our results should directly apply. In this network each actor is a node, and two actors are connected
if they were cast in the same movie during their career.
This network grows continuously by the addition of new
actors (nodes). However,
an important contribution to the network connectivity
comes from another frequent event: a veteran actor plays in a new movie, establishing new
internal links (process (i)) with actors (nodes) with whom
he/she did not play before. Since links are
never eliminated, rewiring is absent $(q=0)$, thus the evolution of this network can
be described by a two parameter model ($p, m$): with
probability $p$ a new actor is introduced, playing
with $m$ other actors, and with probability $1-p$, $m$ new
internal links are generated by established actors playing
in new movies. Naturally, in reality the number of actors cast in a movie varies from one movie to another. It can be shown, however, that fluctuations in $m$ do not change the exponent $\gamma$\cite{tu}. To obtain $p$ and $m$, we fit the
connectivity distribution $P(k)$ obtained for this network
with Eq.$\,$(\ref{power}), obtaining an excellent overlap for
$p=0.937$ and $m=1$ (Fig.$\,$3). The corresponding parameters of Eq.$\,$(\ref{power}) are $\kappa(p,q)=31.68$ and
$\gamma(p,q,m)=3.07$. This indicates that $93.7\%$ of new
links connect existing nodes, and only
$6.3\%$ of links come from new actors  joining the movie
industry. Also, $m=1$ indicates that in average
only one new link is formed with every new movie, supporting the high
interconnectivity of the movie industry: although many actors
collaborate in a movie, most of them already played together in previous movies, and thus links
already exist between them.

{\it Discussion:} While in critical phenomena
power-law scaling is typically associated with universality,
implying that the exponents are independent of the
microscopic details of the model, here we demonstrated that no such universality
exists for scale-free networks, the scaling exponents depending
continuously on the network's parameters. On the other hand,
our results  indicate the existence
of a different criterium for universality based on the
functional form of $P(k)$: Our model predicts the existence of two regimes, the scale-free and the exponential regime. Some of the large networks investigated so far, such as the www or the actor network, are described by scale-free networks\cite{clever,diam,ba,faloutsos}. However, a number of fundamental
network models\cite{er,ws,amaral} lead to $P(k)$ that decays exponentially, indicating the robustness of the exponential regime as well\cite{mechanisms}.

We find that the elementary processes and their frequencies uniquely determine the topology of the system and
the form of $P(k)$.  This suggests that by monitoring their rate we can predict the large scale topology of real networks. Furthermore, our results can be used to
achieve inverse engineering as well: by fitting $P(k)$
obtained for the actor network, we are able to determine the
frequency of the elementary processes. We believe that these
results will help us better understand
the topology, dynamics and evolution of various complex
networks, with potential applications
ranging from biological to social and communication networks.

We wish to acknowledge useful discussions with Y. Tu, H.
Jeong, S. Yook and thank  B. Tjaden for providing us the data
on the actor collaborations. This work was partially supported by the NSF Career Award DMR-9710998.

\begin{figure}[ht] 
\centerline{\hspace{-1cm}\psfig{figure=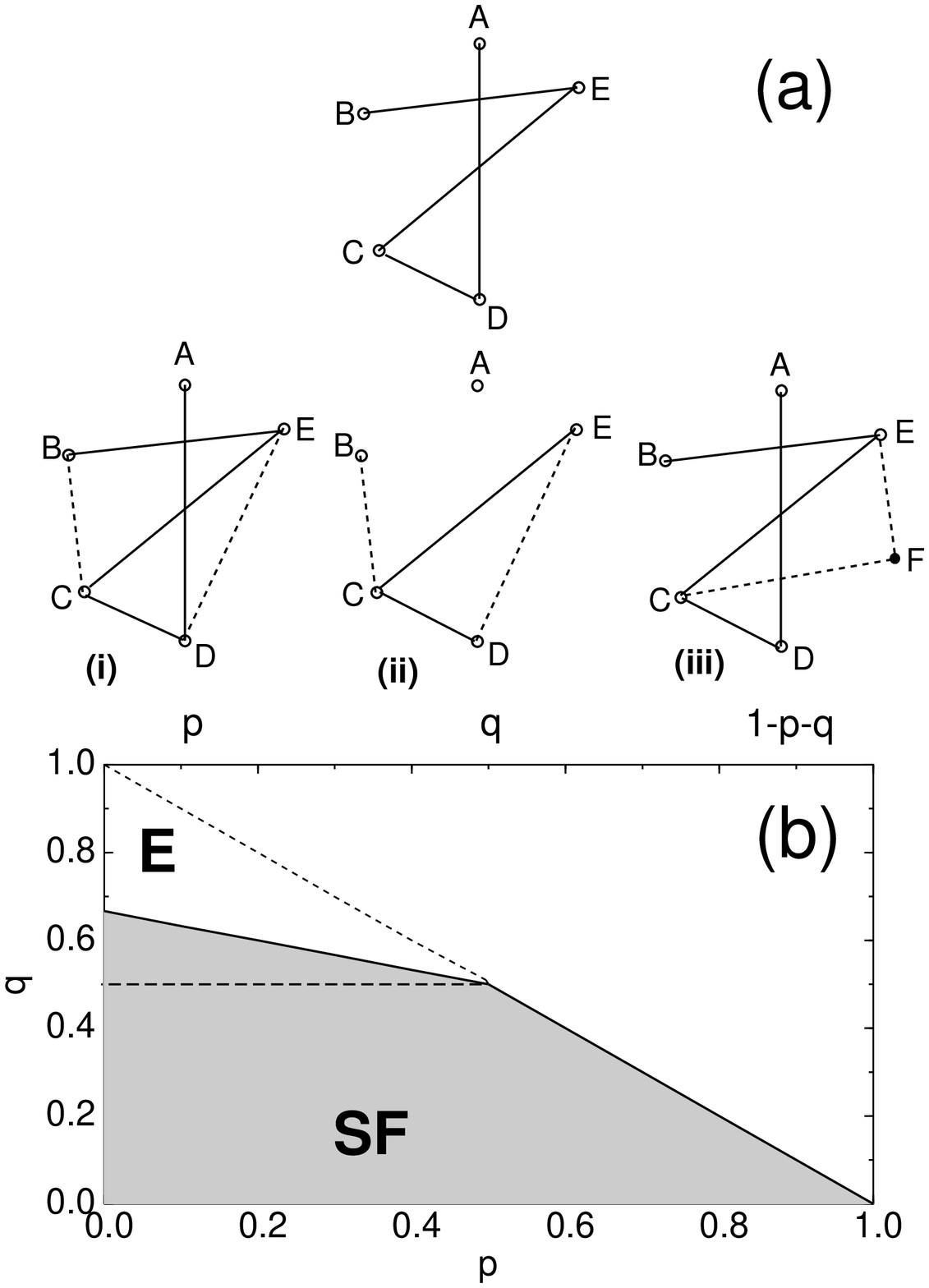,height=5.6in,width=4.5in}}
\renewcommand\baselinestretch{0.9}
\caption{(a) Illustration of the possible elementary processes in the
extended model for $m_0=3$ and $m=2$. At time $t=2$
there are five nodes in the system,
connected by four links (upper panel).
In the next timestep, one of the three possible events can take place: (i) Add
$m=2$ links with probability $p$. One end
of the new link is selected randomly, the other is selected using preferential attachment (Eq.$\,$(\ref{prefer})). The new links are
drawn with dashed lines; (ii) Rewire $m=2$ links with
probability $1-q$. The AD link is disconnected from its A
end and connected preferentially to the highly connected
node E; (iii) Add
a new node (F) and connect it with $m=2$ links to the nodes in the system with probability
$1-p-q$, using preferential attachment.
(b) Phase diagram for the extended model. The scale-free regime (SF) for $m=1$ is shaded, the remaining of the $p+q<1$ domain corresponding to the exponential regime (E). The boundary between E and SF is shown as a dotted line when $m\rightarrow 0$, or as a dashed line when $m\rightarrow\infty$.  
}
\end{figure}

\begin{figure}[ht]
\centerline{\hspace{-3.5cm}\psfig{figure=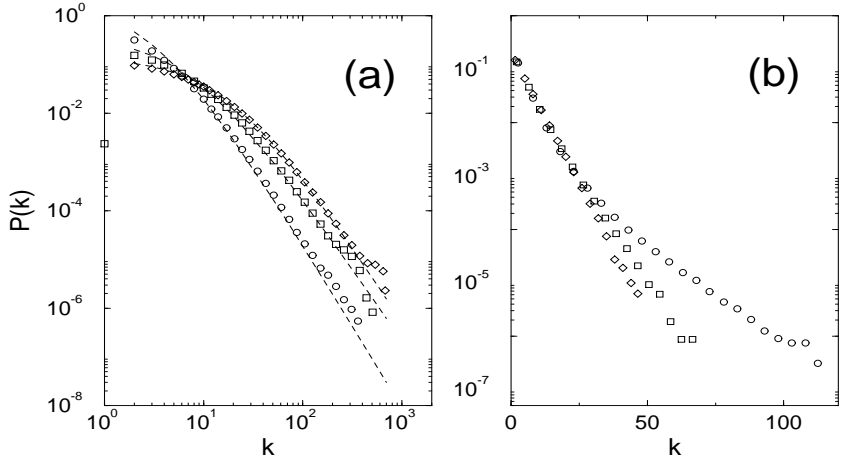,height=3.3in,width=2.2in,angle=-90}}
\renewcommand\baselinestretch{0.9}
\caption{(a) Comparison between the numerical simulations and the prediction of the continuum theory in the scale-free regime. In the simulations we used $t=100,000$, $m_0=m=2$. Circles: $p=0.3$, $q=0$; squares: $p=0.6$, $q=0.1$; diamonds: $p=0.8$, $q=0$. The data were logarithmically binned. The parameter free predictions of Eq.$\,$(\ref{power}) are shown as dashed lines\protect\cite{note}. (b) The numerically obtained $P(k)$ in the exponential regime, shown on a semi-logarithmic plot, indicating the convergence of $P(k)$ to an exponential in the $q\rightarrow 1$ limit. Circles: $p=0$, $q=0.8$; squares: $p=0$, $q=0.95$; diamonds: $p=0$, $q=0.99$. }
\end{figure}

\vspace{1cm}
\begin{figure}[hb]
\centerline{\hspace{-2cm}\psfig{figure=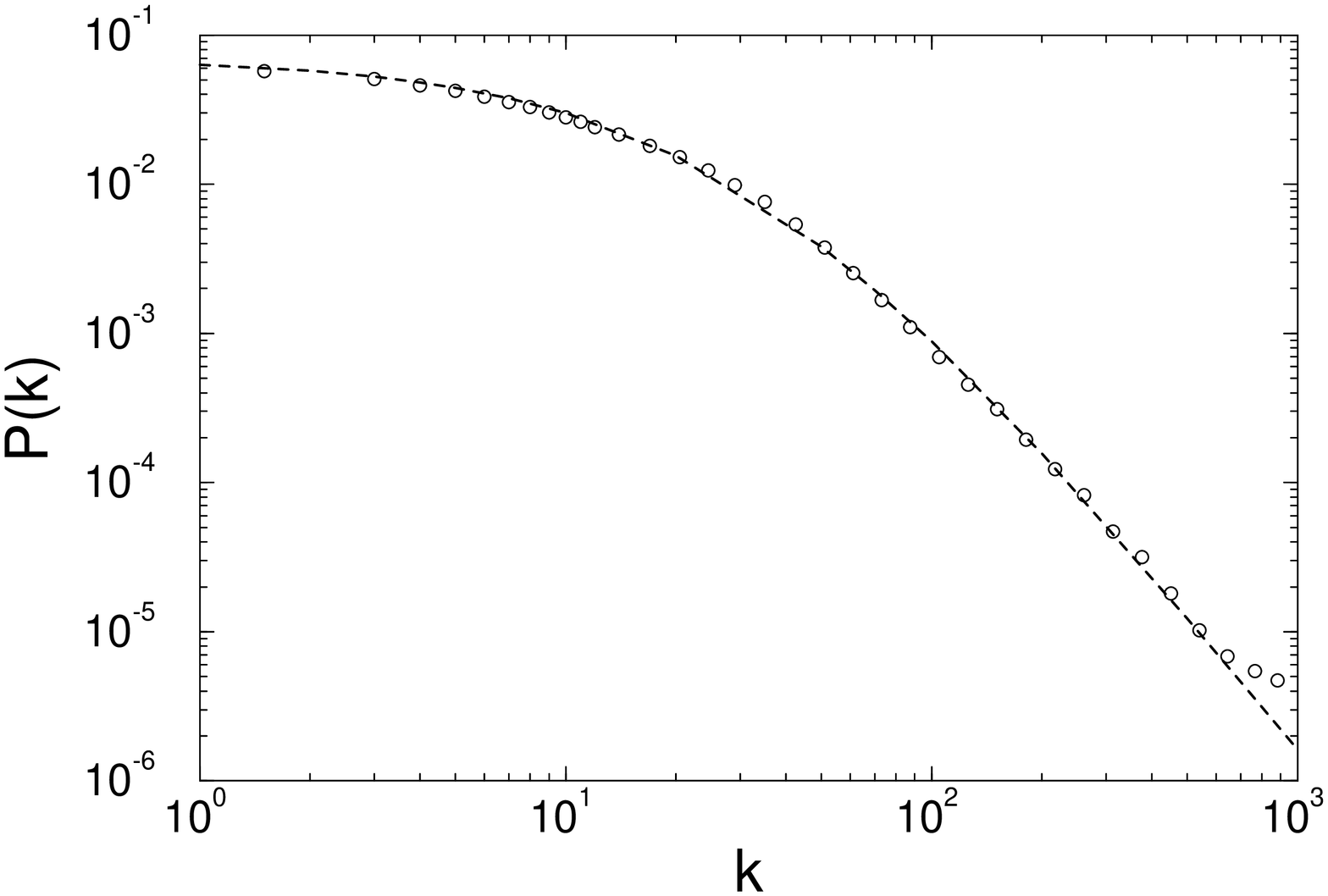,height=3in,width=2in,angle=-90}}
\renewcommand\baselinestretch{0.9}
\caption{The connectivity distribution of movie actors (circles) based on the IMDB database containing $212,250$ actors and $61,085,555$ links. The data were logarithmically binned. The dashed line corresponds to the two parameter fit offered by Eq.$\,$(\ref{power}) with $p=0.937$ and $m=1$\protect\cite{note}.}
\end{figure}


\begin{references}
\bibitem{note}
Note that we measured the consecutive slopes\protect\cite{zipf} $dlog(P(k))/dlog(k)$, further confirming the agreeement between the analytical prediction and the numerical results or actor network distribution.
\end{references}
\end{document}